\documentclass{aastex61}
\usepackage{xspace}
\usepackage{natbib}
\newcommand\aastex{AAS\TeX}

\newcommand\trilegal{TRILEGAL\xspace}
\newcommand{\kk}{{\it{K2}}\xspace}
\newcommand{\kepler}{{\it{Kepler}}\xspace}
\newcommand{\tess}{{\it{TESS}}\xspace}

\shorttitle{\aastex\ \kk Stellar Companions}
\shortauthors{Matson et al.}

\begin{document}

\title{Stellar Companions of Exoplanet Host Stars in K2}

\correspondingauthor{Rachel Matson}
\email{rachel.a.matson@nasa.gov}

\author[0000-0002-0786-7307]{Rachel A. Matson}
\affil{NASA Ames Research Center, 
Moffett Field, CA 94035, USA}

\author{Steve B. Howell}
\affiliation{NASA Ames Research Center, 
Moffett Field, CA 94035, USA}

\author{Elliott P. Horch}
\affiliation{Department of Physics, Southern Connecticut State University, 501 Crescent Street, New Haven, CT 06515, USA}

\author{Mark E. Everett}
\affiliation{National Optical Astronomy Observatory, 950 North Cherry Avenue, Tucson, AZ 85719, USA}

\begin{abstract}

It is well established that 
roughly half of all nearby solar-type stars have at least one companion. Stellar companions can have significant implications for the detection and characterization of exoplanets, including triggering false positives and masking the true radii of planets. Determining the fraction of exoplanet host stars that are also binaries allows us to better determine planetary characteristics as well as establish the relationship between binarity and planet formation. Using high angular resolution speckle imaging, we detect stellar companions within $\sim$\,1 arcsec of \kk planet candidate host stars. Comparing our detected companion rate to TRILEGAL star count simulations and known detection limits of speckle imaging we estimate the binary fraction of \kk planet host stars to be $40-50\%$, similar to that of \kepler exoplanet hosts and field stars.

\end{abstract}


\section{Introduction} \label{sec:intro}

Binary stars have long been important tools in astrophysics, providing constraints on stellar structure and evolution through empirically determined parameters and by informing our understanding of star formation mechanisms and outcomes. Understanding the frequency of binary star systems also has significant implications for star and planet formation processes, and how they are affected by the presence of a companion \citep{Duchene:araa2013a}. The frequency of multiple star systems is a function of primary star mass, with approximately $46\%$ of solar-type stars having at least one companion \citep{Raghavan:apjs2010a}. This is especially relevant for exoplanets studies, as nearly half of all sun-like stars form with at least one companion and we do not yet understand the impact of stellar companions on planet formation \citep{Kraus:aj2016a}.

From a theoretical standpoint, binary companions should have an adverse dynamical influence on planet formation processes, including perturbing and truncating protoplanetary disks \citep{Jang-Condell:apj2015a}, gravitationally exciting planetesimals causing collisional destruction \citep{Rafikov:apj2015a, Rafikov:apj2015b}, and causing dynamical interactions that can scatter or eject planets that have formed \citep{Haghighipour:apj2006a}. However, discoveries of numerous planets in binary systems, including binaries with small separations and circumbinary planets, are raising questions regarding the formation, long-term stability, and habitability of planets in multiple star systems \citep{Thebault:2015a}. 

The \kepler mission, in particular, has discovered nearly 7000 confirmed and candidate exoplanets with a surprising diversity of properties and system architectures. Unlike many exoplanet surveys, \kepler was largely unbiased toward binary stars and therefore provides the opportunity to study the influence of stellar companions on planet occurrence. Studies of \kepler exoplanet host stars with stellar companions by \citet{Horch:apj2014a} and \citet{Deacon:mnras2016a} found no difference in stellar multiplicity between exoplanet host stars and nearby field stars. However, studies by \citet{Wang:apj2014a, Wang:apj2014b} and \citet{Kraus:aj2016a} have found evidence for fewer close binary companions ($\lesssim 100$\,AU) in \kepler exoplanet host stars. 

To examine this discrepancy further and investigate a more varied sample of exoplanet host stars, we evaluate stellar companions of exoplanet candidate host stars discovered with the repurposed \kepler mission, \kk. \kk observes fields along the ecliptic in campaigns of approximately 80 days \citep{Howell:pasp2014a}, and thus far has discovered more than 500 exoplanet candidates.

The \kk targets include bright solar-like stars and low-mass stars distributed across a wide range of galactic latitudes, many of which are suitable for follow-up characterization and atmospheric studies, serving as a precursor for the wide-field {\it{Transiting Exoplanet Survey Satellite}} (\tess) mission. Understanding the role of stellar multiplicity in such exoplanet systems is vital for accurately characterizing  exoplanets and identifying systems for follow-up studies such as transit spectroscopy with the {\it{James Webb Space Telescope}} ({\it{JWST}}) and direct imaging with the {\it{Wide-Field Infrared Survey Telescope}} ({\it{WFIRST}}). 

In addition, the pixel scale of \kepler/\kk ($\sim$\,$4''$), and even larger pixel scale of \tess ($\sim$\,$21''$), can result in flux from background objects and companion stars being blended with the host star. If the photometric aperture contains flux from nearby stars the measured transit depth will be smaller, causing the radius of the planet to be underestimated \citep{Ciardi:apj2015a}. Alternately, if the neighboring star is an eclipsing binary, the eclipses may be diluted such that they mimic transit-like signals in the light curve of the target star, producing a false positive \citep{Brown:apjl2003a}. Therefore, understanding the frequency of bound and line-of-sight companions plays a vital role in our detection and characterization of exoplanets. 

To confirm the validity of transiting exoplanet candidates, follow-up observations such as high-resolution imaging are used to detect nearby stellar companions and determine whether contaminating flux is responsible for transit-like signals in the light curve. Here we focus on the technique of speckle imaging, which can deliver diffraction-limited images of exoplanet host stars over a range of magnitudes (\kepler magnitude, $K_p \sim 8 - 17$) with relatively high dynamic range, thereby measuring the brightness and location of nearby companions within $<0.1-1''$ and up to 6 magnitudes fainter than the exoplanet host star \citep{Howell:aj2011a}. Such observations have been used to validate exoplanets from \kepler and \kk \citep[e.g.,][]{Horch:aj2012a, Everett:aj2015a, Crossfield:apjs2016a, Dressing:aj2017a}, as well as catalog stellar companions \citep{Furlan:aj2017a} and account for their effects on planetary properties and occurrence rates \citep{Ciardi:apj2015a, Hirsch:aj2017a, Furlan:aj2017b}.

In this paper we identify a subset of exoplanet candidate host stars detected by \kk and observed with speckle imaging in order to examine the binarity of \kk exoplanet hosts by comparing the observed companion fraction to that of a simulated stellar population. In Section \ref{sec:obs} we describe the speckle observations collected for this work and the properties of the observed exoplanet host stars. Section \ref{sec:methods} describes our simulated stellar population, including how we assign stellar companions to the simulated stars and the detection limits used to determine which stars and companions are observable via speckle imaging. We then present the observed and simulated companion fractions and compare the results in Section \ref{sec:results}. Lastly, in Section \ref{sec:compare} we compare our results for \kk to the \kepler results of \citet{Horch:apj2014a} and examine our sample for suppression of close binary companions.

\section{Sample} \label{sec:obs}

\subsection{Observations} \label{subsec:obs}

The speckle data examined in this work were taken between 2015 September and 2016 June for high-resolution follow-up imaging of planet-candidate host stars detected with \kk through campaign seven. Although the goal was to observe as many planet hosts as possible, systems with planets estimated to be 3\,$R_{\earth}$ or less were given higher priority as they are nearly impossible to validate with other methods \citep{Howell:apjl2016a}. Observations were conducted with the Differential Speckle Survey Instrument \citep[DSSI;][]{Horch:aj2009a} using the WIYN 3.5\,m Telescope at Kitt Peak National Observatory, the Gemini North 8.1\,m Telescope on Mauna Kea, and the Gemini South 8.1\,m Telescope on Cerro Pachon. Most targets were observed only once, however, $\sim$\,$10\%$ were observed $2-3$ times and are listed as separate observations in Table \ref{tab:dataobs}. More details of the speckle observing and data reduction procedures can be found in e.g.,~\citet{Howell:aj2011a}, \citet{Horch:aj2011b}, and \citet{Horch:aj2012a}. 

DSSI consists of two electron-multiplying CCDs that record speckles in different filters simultaneously, resulting in two diffraction-limited reconstructed images. The reconstructed images and 5-$\sigma$ detection limit curves for both filters are available for all \kk observations at the Exoplanet Follow-up Observing Program (ExoFOP) website\footnote{\url{https://exofop.ipac.caltech.edu/}}. Since 2008, filters centered at 562\,nm, 692\,nm, and 880\,nm have been used on DSSI. The 880\,nm filter (54\,nm wide) usually provides a larger dynamic range, making it sensitive to fainter companions, and was used to observe all \kk planet host candidates in this sample at least once. As not all stars were observed in any other filter, we use the 880\,nm data for this study and apply the corresponding filter transmission curve to the simulated data when comparing detection rates (see Section \ref{sec:methods}). Table \ref{tab:dataobs} shows the number of speckle observations by filter and observatory, as well as the number of unique stars and companions observed with DSSI. Duplicate observations of four systems with detected companions were conducted at both WIYN and Gemini. A companion was detected at both telescopes (WIYN and Gemini South) for two of the systems and detected only at Gemini in the other two, which is reflected in the number of companions detected at each observatory vs.~the total number detected (29) in Table \ref{tab:dataobs}.

\begin{deluxetable}{lCCcc}[t]
\tablecaption{Observed \kk Exoplanet Host Stars \label{tab:dataobs}}
\tablecolumns{5}
\tablenum{1}
\tablewidth{0pt}
\tablehead{
\colhead{} &
\colhead{Gemini-N} &
\colhead{Gemini-S} &
\colhead{\phn WIYN \phn} &
\colhead{Total}
}
\startdata
562\,nm Observations & 0 & 11 & 0 & 11 \\
692\,nm Observations & 34 & 57 & 127 & 218 \\
880\,nm Observations & 34 & 68 & 127 & 229 \\
Total Observations & 68 & 136 & 254 & 458 \\
\hline
Companions Detected (880\,nm) & 10 & 13 & 8 & 29\\
Unique Stars Observed & 34 & 68 & 127 & 206 \\
Average \kepler Magnitude & 12.7 & 12.6 & 12.3 & 12.5 \\
\enddata
\end{deluxetable}


\begin{figure}[h]
\includegraphics[scale=0.5]{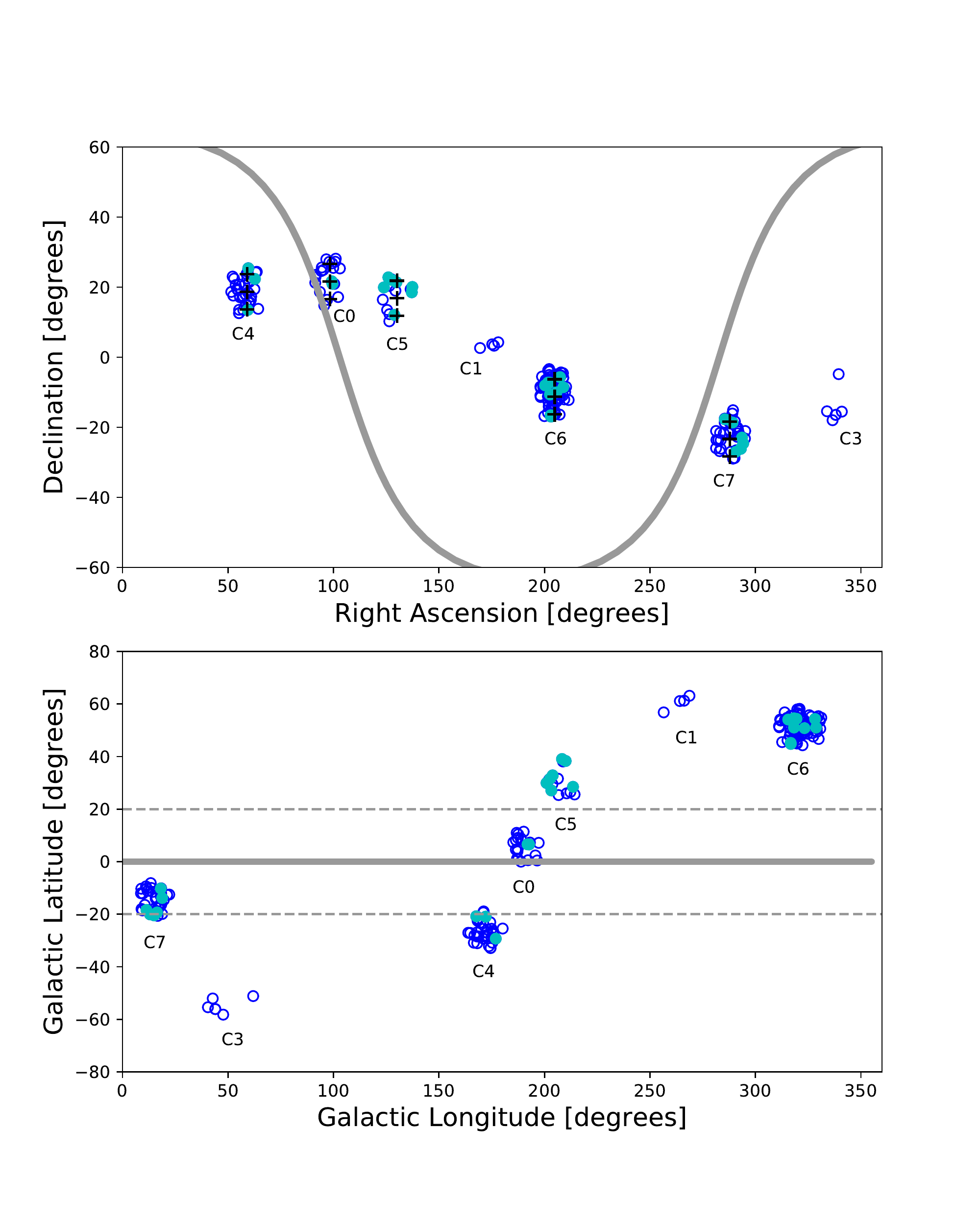}
\centering
\caption{Locations of planet-candidate host stars identified by \kk and observed with speckle imaging in equatorial (top) and Galactic coordinates (bottom). Filled cyan circles indicate stars with a detected companion, while open blue circles indicate no companion was detected. The Galactic plane is shown by the solid gray line and \kk campaign numbers are noted for all observed campaigns. Black plus symbols in the top plot indicate the coordinates of the \trilegal pointings (see Section \ref{sec:methods}). In the bottom plot dashed gray lines highlight the region $\pm 20 \degr$ the Galactic plane. 
\label{fig:coords}}
\end{figure}

The equatorial and galactic coordinates of all \kk stars observed with DSSI are plotted in Figure \ref{fig:coords}. Filled cyan dots represent systems with detected companions, while open blue dots show observed systems where no companions were detected. The solid gray line depicts the location of the Galactic plane in both plots, with the dashed lines in the bottom plot highlighting the region within $\pm 20 \degr$ of the Galactic plane. Individual \kk campaign fields are labelled in black.

For the 206 unique \kk planet-candidate host stars observed with DSSI, 29 companions were detected with the 880nm filter. The Ecliptic Plane Input Catalog \citep[EPIC;][]{Huber:apjs2016a} ID, \kk Campaign number, date of observation, and relevant measurements for the detected companion stars are given in Table \ref{tab:companions}. The central wavelength and width of the filter used for each observation are listed as $\lambda\,(\Delta \lambda)$ in nm, followed by the separation ($\rho$) of the two stars in arcseconds, the position angle ($\theta$) of the secondary star relative to the primary in degrees from north, the magnitude difference ($\Delta$m) of the two stars, and the 5-$\sigma$ detection limits at 0.2'', 0.5'', and 1.2'' from the primary star. The accuracy and precision of the astrometry is determined by comparing measured position angles and separations to ephemeris positions of objects with known orbits \cite[e.g.,][]{Horch:aj2011a} or binaries with multiple observations for \kepler/\kk targets \citep{Horch:aj2012a}. For binaries observed with DSSI at WIYN, the standard deviation between the speckle measurements and those of known binaries is less than $0.5 \degr$ in position angle and between $1-2$ mas in separation \citep{Horch:aj2011a}, with analogous results obtained for \kepler targets observed at Gemini \cite[$\theta \sim 0.5 \degr$ and $\rho \sim 1-3$ mas;][]{Horch:aj2012a}. Similarly, the photometric precision is determined through comparisons with {\it{Hipparcos}} magnitudes or other large-aperture speckle observations taken in similar filters \cite[e.g.,][]{Horch:aj2011a, Horch:aj2012a}, leading to an estimate of DSSI's photometric precision of $\lesssim$\,0.15 mag. The method used to determine the 5-$\sigma$ detection limits is outlined in Section \ref{sec:methods}, or see \cite{Horch:aj2011a} and \citet{Howell:aj2011a} for more detail. All data summarized in Table \ref{tab:companions} are available at the ExoFOP website, including reconstructed images and full detection limit curves.

\startlongtable
\begin{deluxetable}{cccccCCCccc}
\tablecaption{Companion Detections \label{tab:companions}}
\tabletypesize{\small}
\tablecolumns{11}
\tablenum{2}
\tablewidth{0pt}
\tablehead{
\colhead{EPIC ID} &
\colhead{Campaign} &
\colhead{Telescope} &
\colhead{Date} &
\colhead{$\lambda$ ($\Delta \lambda$) } &
\colhead{Separation} &
\colhead{Position Angle} &
\colhead{$\Delta$m} &
\multicolumn{3}{c}{Limiting $\Delta$m at:}
\vspace{-2pt} \\
\colhead{} &
\colhead{No.} &
\colhead{} &
\colhead{(MJD)} &
\colhead{(nm)} &
\colhead{('')} &
\colhead{($\degr$)} &
\colhead{(mag)} &
\colhead{0.2''} &
\colhead{0.5''} &
\colhead{1.2''}
}
\startdata
202059377 &	0 & 	WIYN 3.5m	& 	57319.378 & 692 (47) &	0.407   &	260.100	& 0.64 	  & 2.607 & 3.394 & 4.320 \\
          & 0 & 	WIYN 3.5m	& 	57319.378 & 880 (54) &	0.403   &	260.300	& 0.67 	  & 2.643 & 3.129 & 3.478 \\
202072596 &	0 & 	WIYN 3.5m	& 	57319.406 & 692 (47) &	0.879   &	306.800	& 1.38 	  & 3.456 & 3.411 & 3.766 \\
          & 0 & 	WIYN 3.5m	& 	57319.406 & 880 (54) &	0.873   &	306.400	& 1.38 	  & 2.524 & 3.247 & 3.408 \\
210401157 & 4 & 	Gemini-N 8m & 	57402.356 & 692 (47) &	0.470   &	161.393	& 2.47 	  & 4.832 & 5.740 & 6.972 \\
          & 4 & 	Gemini-N 8m & 	57402.356 & 880 (54) &	0.488  &	162.529	& 2.30 	  & 4.342 & 5.778 & 7.216 \\
210958990 & 4 & 	Gemini-N 8m & 	57402.389 & 692 (47) &	1.650   &	135.316	& 2.71 	  & 4.445 & 5.313 & 5.717 \\
          & 4 & 	Gemini-N 8m & 	57402.389 & 880 (54) &	1.793   &	133.880	& 2.38 	  & 4.819 & 5.510 & 6.183 \\
211147528 & 4 & 	Gemini-N 8m & 	57402.361 & 692 (47) &	\nodata	&   \nodata & \nodata & 4.511 & 5.924 & 6.933 \\
          & 4 & 	Gemini-N 8m & 	57402.361 & 880 (54) &	1.338   &	130.724   & 7.99 	  & 4.505 & 5.889 & 6.918 \\
          & 4 & 	WIYN 3.5m	& 	57322.390 & 692 (47) &	\nodata &     \nodata & \nodata & 2.900 & 3.968 & 4.153 \\
          & 4 & 	WIYN 3.5m	& 	57322.390 & 880 (54) &	\nodata &     \nodata & \nodata & 2.846 & 3.260 & 3.516 \\
211428897 & 5 & 	Gemini-N 8m & 	57402.531 & 692 (47) &	1.086   &	43.876	& 1.81 	  & 4.145 & 4.446 & 4.496 \\
          & 5 & 	Gemini-N 8m & 	57402.531 & 880 (54) &	1.119   &	43.694	& 1.15 	  & 4.285 & 4.873 & 5.061 \\
211886472 & 5 & 	Gemini-N 8m & 	57404.462 & 692 (47) &	0.323   &	278.86	& 4.02 	  & 5.463 & 6.952 & 8.691 \\
          & 5 & 	Gemini-N 8m & 	57404.462 & 880 (54) &	0.326   &	279.74	& 3.77 	  & 4.660 & 5.204 & 5.361 \\
211978865 & 5 & 	Gemini-N 8m & 	57400.492 & 692 (47) &	1.068   &	28.379	& 3.44 	  & 4.733 & 4.932 & 5.063 \\
          & 5 & 	Gemini-N 8m & 	57400.492 & 880 (54) &	1.137   &	27.928	& 3.20 	  & 3.717 & 3.984 & 4.330 \\
211987231 & 5 & 	Gemini-N 8m & 	57403.438 & 692 (47) &	0.906   &	239.035	& 1.67 	  & 5.452 & 6.254 & 6.970 \\
          & 5 & 	Gemini-N 8m & 	57403.438 & 880 (54) &	0.924   &	238.772	& 1.46 	  & 5.154 & 5.980 & 6.353 \\
212066407 & 5 & 	Gemini-N 8m & 	57401.490 & 692 (47) &	\nodata &     \nodata & \nodata & 4.572 & 5.030 & 5.296 \\
          & 5 & 	Gemini-N 8m & 	57401.490 & 880 (54) &	0.219   &	183.516	& 4.00 	  & 4.222 & 5.182 & 5.641 \\
          & 5 & 	Gemini-N 8m & 	57403.460 & 692 (47) &	0.209   &	184.181	& 5.32 	  & 5.240 & 6.260 & 7.319 \\
          & 5 & 	Gemini-N 8m & 	57403.460 & 880 (54) &	0.222   &	183.596	& 4.06 	  & 4.707 & 5.981 & 6.640 \\
212099230 & 5 & 	Gemini-N 8m & 	57401.499 & 692 (47) &	\nodata &     \nodata & \nodata & 4.966 & 5.643 & 6.029 \\
          & 5 & 	Gemini-N 8m & 	57401.499 & 880 (54) &	0.105   &	150.888	& 3.14 	  & 4.786 & 6.097 & 6.970 \\
          & 5 & 	Gemini-N 8m & 	57404.448 & 692 (47) &	0.102   &	148.778	& 3.95 	  & 4.823 & 6.923 & 9.294 \\
          & 5 & 	Gemini-N 8m & 	57404.448 & 880 (54) &	0.105   &	150.325	& 3.19 	  & 4.654 & 5.429 & 5.702 \\
212138198 & 5 & 	Gemini-N 8m & 	57403.451 & 692 (47) &	0.250   &	36.936    & 2.06 	  & 5.013 & 5.903 & 6.653 \\
          & 5 & 	Gemini-N 8m & 	57403.451 & 880 (54) &	0.258   &	36.754    & 1.61 	  & 4.483 & 5.401 & 5.910 \\
212303338 & 6 & 	WIYN 3.5m	& 	57496.227 & 692 (47) &	0.098   &	151.199	& 2.45 	  & 3.085 & 3.524 & 3.710 \\
          & 6 & 	WIYN 3.5m	& 	57496.227 & 880 (54) &	0.113   &	153.422	& 1.99 	  & 4.249 & 4.356 & 4.535 \\
212315941 & 6 & 	Gemini-S 8m & 	57562.049 & 692 (47) &	0.064   &	86.159	& 1.12 	  & 4.588 & 4.952 & 5.057 \\
          & 6 & 	Gemini-S 8m & 	57562.049 & 880 (54) &	0.057   &	90.322	& 1.29 	  & 4.132 & 4.704 & 4.782 \\
212534729 & 6 & 	WIYN 3.5m	& 	57502.303 & 692 (47) &	0.151   &	292.368	& 0.63 	  & 3.138 & 3.059 & 3.186 \\
          & 6 & 	WIYN 3.5m	& 	57502.303 & 880 (54) &	0.171   &	288.671	& 1.23 	  & 2.773 & 2.956 & 3.195 \\
212565386 & 6 & 	Gemini-S 8m & 	57561.074 & 692 (47) &	0.140   &	251.506	& 1.34 	  & 4.708 & 4.923 & 5.034 \\
          & 6 & 	Gemini-S 8m & 	57561.074 & 880 (54) &	0.140   &	250.912	& 0.90 	  & 4.497 & 4.723 & 4.961 \\
212577658\tablenotemark{a}& 6 & 	WIYN 3.5m	& 	57496.312 & 692 (47) &	1.810   &	174.510	& 0.01 	  & 3.091 & 3.444 & 3.628 \\
          & 6 & 	WIYN 3.5m	& 	57496.312 & 880 (54) &	\nodata &     \nodata & \nodata & 3.390 & 3.451 & 3.521 \\
212628098 & 6 & 	Gemini-S 8m & 	57561.047 & 692 (47) &	\nodata &     \nodata & \nodata & 4.756 & 5.086 & 5.532 \\
          & 6 & 	Gemini-S 8m & 	57561.047 & 880 (54) &	1.254   &	161.736	& 3.82 	  & 4.432 & 5.149 & 5.304 \\
          & 6 & 	WIYN 3.5m	& 	57498.237 & 692 (47) &	\nodata &     \nodata & \nodata & 2.573 & 2.688 & 2.930 \\
          & 6 & 	WIYN 3.5m	& 	57498.237 & 880 (54) &	\nodata &     \nodata & \nodata & 2.423 & 2.538 & 2.733 \\
212651213 & 6 & 	WIYN 3.5m	& 	57495.327 & 692 (47) &	0.114   &	68.641	& 0.55 	  & 3.576 & 3.757 & 4.101 \\
          & 6 & 	WIYN 3.5m	& 	57495.327 & 880 (54) &	0.110   &	68.442	& 0.35 	  & 3.080 & 3.624 & 3.777 \\
212679181 & 6 & 	Gemini-S 8m & 	57560.083 & 692 (47) &	1.245   &	30.015	& 1.07 	  & 4.526 & 4.798 & 4.846 \\
          & 6 & 	Gemini-S 8m & 	57560.083 & 880 (54) &	1.250   &	29.649	& 1.12 	  & 4.371 & 4.734 & 4.944 \\
          & 6 & 	WIYN 3.5m	& 	57495.319 & 692 (47) &	1.478   &	29.730	& 1.48 	  & 3.068 & 3.520 & 3.710 \\
          & 6 & 	WIYN 3.5m	& 	57495.319 & 880 (54) &	1.457   &	30.338	& 1.15 	  & 3.362 & 3.533 & 3.774 \\
212679798 & 6 & 	Gemini-S 8m & 	57561.064 & 692 (47) &	\nodata &     \nodata & \nodata & 4.994 & 4.978 & 5.144 \\
          & 6 & 	Gemini-S 8m & 	57561.064 & 880 (54) &	0.112   &	339.227	& 2.61 	  & 4.492 & 4.834 & 4.917 \\
212703473 & 6 & 	WIYN 3.5m	& 	57495.267 & 692 (47) &	0.248   &	307.739	& 1.14 	  & 3.559 & 4.114 & 4.283 \\
          & 6 & 	WIYN 3.5m	& 	57495.267 & 880 (54) &	0.245   &	307.482	& 0.94 	  & 3.200 & 3.291 & 3.457 \\
212773309 & 6 & 	Gemini-S 8m & 	57560.063 & 692 (47) &	1.010   &	245.800	& 2.80 	  & 4.904 & 5.861 & 6.441 \\
          & 6 & 	Gemini-S 8m & 	57560.063 & 880 (54) &	1.009   &	245.354	& 1.99 	  & 4.894 & 5.858 & 6.680 \\
          & 6 & 	WIYN 3.5m	& 	57502.363 & 692 (47) &	1.197   &	245.652	& 2.83 	  & 3.434 & 3.861 & 4.193 \\
          & 6 & 	WIYN 3.5m	& 	57502.363 & 880 (54) &	1.180   &	246.190	& 2.04 	  & 3.106 & 3.399 & 3.613 \\
213563657 & 7 & 	Gemini-S 8m & 	57560.294 & 692 (47) &	0.933   &	162.696	& 2.31 	  & 4.681 & 4.787 & 4.904 \\
          & 7 & 	Gemini-S 8m & 	57560.294 & 880 (54) &	0.963   &	350.970	& 3.28 	  & 4.084 & 4.409 & 4.538 \\
213919915 & 7 & 	Gemini-S 8m & 	57559.233 & 692 (47) &	1.084   &	18.515	& 0.89 	  & 4.341 & 4.880 & 5.109 \\
          & 7 & 	Gemini-S 8m & 	57559.233 & 880 (54) &	1.090   &	18.226	& 0.96 	  & 4.119 & 4.814 & 5.373 \\
213920015 & 7 & 	Gemini-S 8m & 	57568.254 & 692 (47) &	1.081   &	18.649	& 0.85 	  & 4.132 & 4.996 & 5.440 \\
          & 7 & 	Gemini-S 8m & 	57568.254 & 880 (54) &	1.092   &	18.274	& 0.87 	  & 4.532 & 5.123 & 5.709 \\
214889247 & 7 & 	Gemini-S 8m & 	57567.328 & 562 (43) &	0.270   &	62.194	& 7.01 	  & 5.058 & 6.173 & 7.168 \\
          & 7 & 	Gemini-S 8m & 	57567.328 & 880 (54) &	0.245   &	60.756	& 4.60 	  & 5.227 & 6.503 & 7.661 \\
216050437 & 7 & 	Gemini-S 8m & 	57567.319 & 562 (43) &	0.086   &	148.263	& 0.20 	  & 4.565 & 5.270 & 5.735 \\
          & 7 & 	Gemini-S 8m & 	57567.319 & 880 (54) &	0.085   &	148.409	& 0.44 	  & 4.361 & 5.080 & 5.656 \\
218131080 & 7 & 	Gemini-S 8m & 	57568.329 & 692 (47) &	0.208   &	189.384	& 2.42 	  & 4.689 & 4.882 & 5.008 \\
          & 7 & 	Gemini-S 8m & 	57568.329 & 880 (54) &	0.196   &	187.294	& 2.55 	  & 4.626 & 4.740 & 4.847 \\
218711655 & 7 & 	Gemini-S 8m & 	57567.368 & 562 (43) &	0.035   &	297.399	& 1.01 	  & 5.110 & 5.591 & 6.148 \\
          & 7 & 	Gemini-S 8m & 	57567.368 & 880 (54) &	0.026   &	297.887	& 1.44 	  & 4.696 & 5.811 & 6.687 \\
\enddata
\tablenotetext{a}{Not included in the observed companion fraction as the companion was not identified at 880nm, but listed here for completeness.}
\end{deluxetable}

Using the WIYN 3.5\,m Telescope, stellar companions with magnitude differences of $\Delta m \sim 3-4$ can be detected to separations of $\sim$\,$0.05"$ around $V \lesssim 14.5$ magnitude stars. Due to the greater light gathering power and smaller diffraction limit of the 8.1\,m Gemini Telescopes, however, companions can be detected at separations of 0.027" and $\Delta m \sim 5-6$ for $V \lesssim 16.5$ magnitude stars. Because of the different detection limits of these telescopes, we consider the companions detected at each telescope separately and compare them to simulated stellar populations with the appropriate detection limits.

\subsection{Stellar Parameters} \label{subsec:stparms}

\begin{figure}[t!]
\includegraphics[scale=0.48]{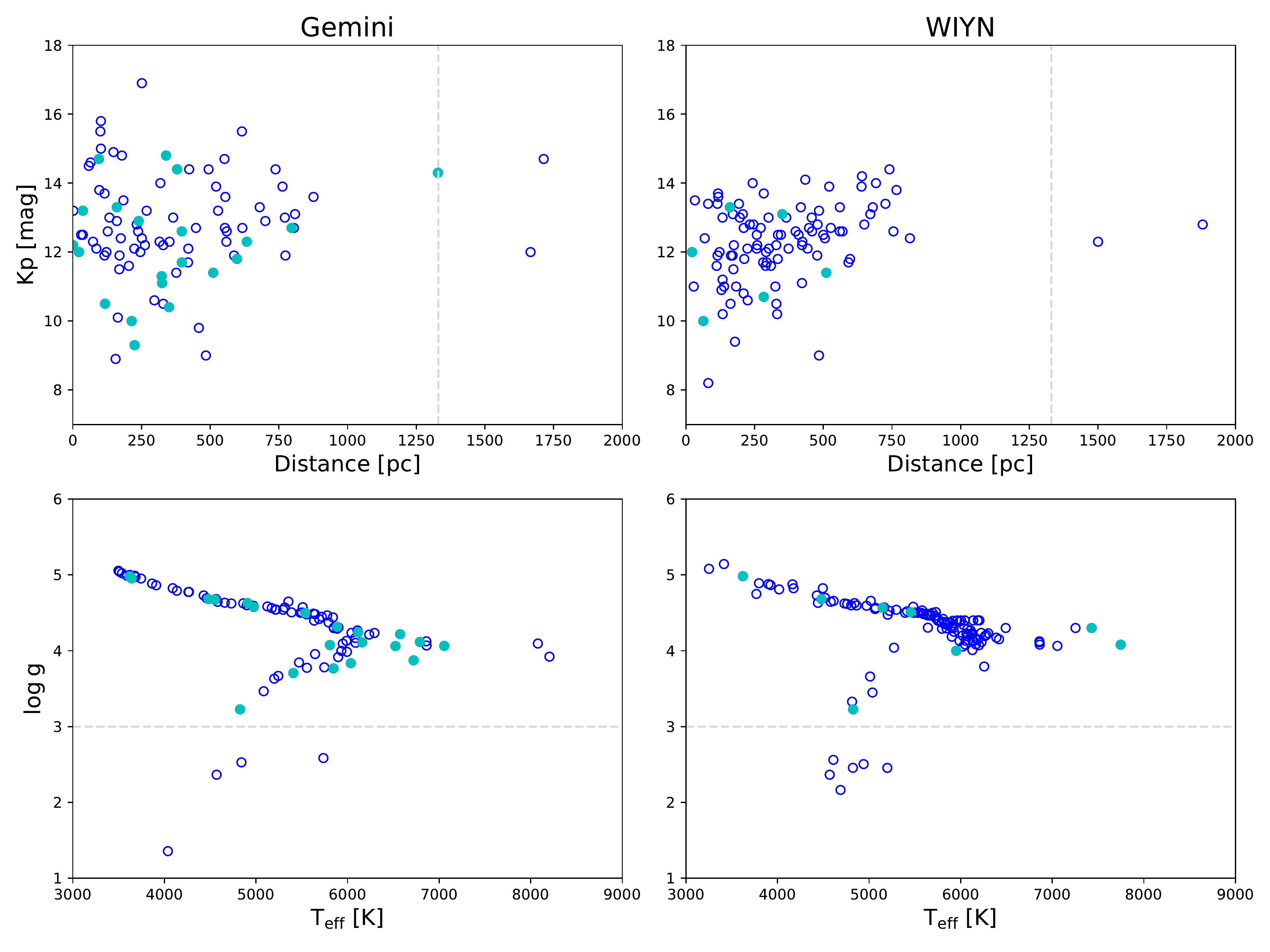}
\centering
\caption{Parameters of planet-candidate host stars identified by \kk and observed with speckle imaging. Filled cyan circles indicate stars with a detected companion, while open blue circles indicate no companion was detected. Plots on the left show stars observed at Gemini North and Gemini South, while stars observed at WIYN are plotted on the right. Gray dashed lines show where cuts were made to the observed data ($d > 1300$\,pc, $\log g < 3.0$) and \trilegal simulations (see Section \ref{sec:methods}) to maintain a distance-limited sample of primarily dwarf stars. 
\label{fig:parms}}
\end{figure}

The stellar properties of the \kk exoplanet host stars observed at Gemini (North and South; left) and WIYN (right) are plotted in Figure \ref{fig:parms}. Filled cyan dots represent systems with detected companions, while open blue dots show systems where no companions were detected. The top panels show the \kepler magnitude of the target stars as a function of estimated distance, with surface gravity as a function of effective temperature plotted in the bottom panels. Effective temperatures, surface gravities, and distances, when available, were adopted from the EPIC as described in \citet{Huber:apjs2016a}. For stars observed during Campaign 0 (C0), which do not appear in the EPIC, we adopt the effective temperatures reported by \cite{Vanderburg:apjs2016a} and estimate the surface gravities based on the physical parameters of dwarf stars in \cite{Gray:2008a} or adopt the effective temperatures and surface gravities of \cite{Crossfield:apjs2016a}. No distance estimates are available for the C0 stars and they are omitted from the top panels of Figure \ref{fig:parms}. In addition, there are six systems with no available stellar parameters that are also omitted from Figure \ref{fig:parms}. The plotted distances and stellar properties do not account for any known or unknown companions, which would affect the estimated distances and stellar properties to some degree depending on the magnitude difference of the components, but are used to show the distribution of the \kk sample and general agreement with the simulated results.

\begin{figure}[t!]
\includegraphics[scale=0.48]{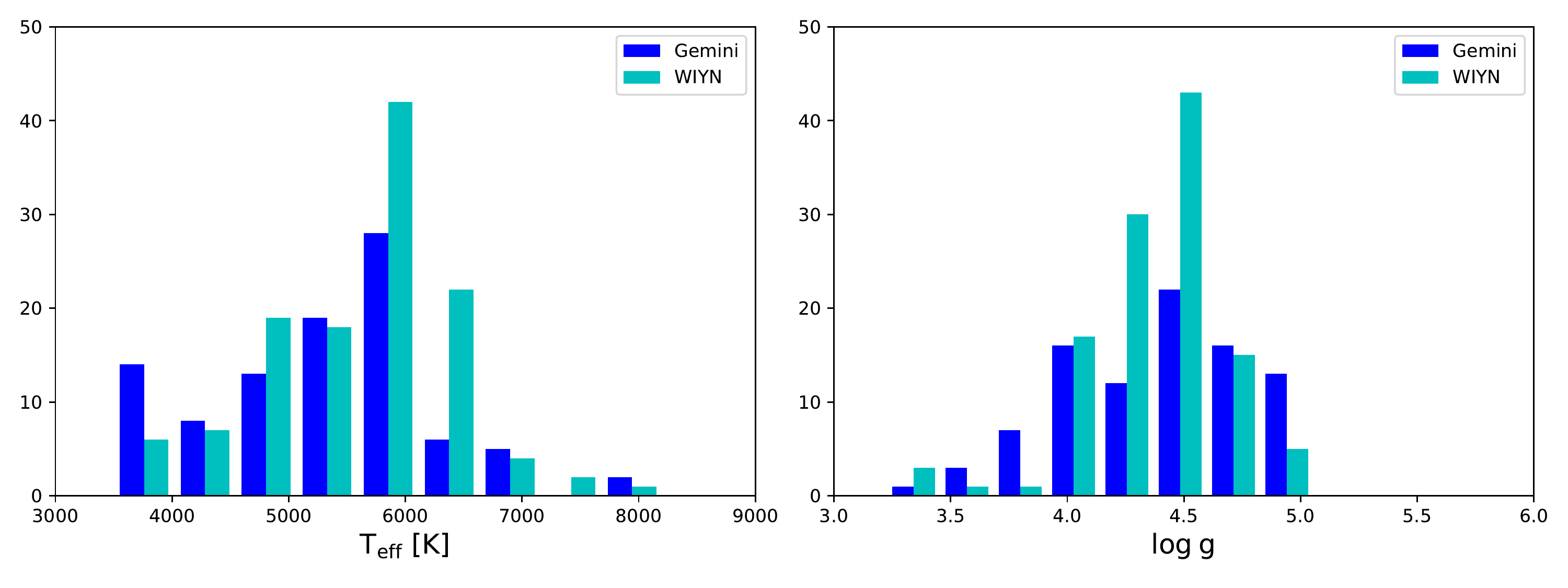}
\centering
\caption{Histograms of effective temperature (left) and surface gravity (right) for \kk exoplanet host stars observed at Gemini (blue) and WIYN (cyan). 
\label{fig:DSSIhist}}
\end{figure}

Histograms of the effective temperature and surface gravity distributions for \kk exoplanet host stars observed at Gemini and WIYN are shown in Figure \ref{fig:DSSIhist}. When compared to the distribution of stellar properties for \kepler exoplanet host stars in \citet[][see their Figure 7]{Horch:apj2014a}, the \kk host star distribution peaks at approximately the same temperature ($\sim 6000$\,K) but has a higher fraction of cooler (fainter) stars. The $\log g$ distributions in both samples appear similar.

Figures \ref{fig:parms} and \ref{fig:DSSIhist} show that our sample spans a range in effective temperature from $\sim$\,3000 $-$ 8000\,K, and is dominated by dwarfs that have near-solar values with estimated distances less than approximately 900pc. This sample is far from uniform or complete as \kk targets are chosen from hundreds of community proposals with varying selection criteria, stellar properties, and science goals. In addition, our speckle observing strategy prioritized stars with small exoplanet candidates within our magnitude range. However, in order to mimic a distance-limited sample of primarily solar-type stars that we can compare to a simulated stellar population and apply the binary statistics of \cite{Raghavan:apjs2010a} to, as well as enable direct comparisons with the binary fraction of \kepler host stars derived by \cite{Horch:apj2014a}, we omit systems with $d > 1300$\,pc and $\log g < 3.0$. Gray dashed lines illustrate the cuts in Figure \ref{fig:parms}. To some extent, K2 is also a magnitude-limited sample, which tend to be overpopulated with binaries due to the excess flux from spatially unresolved companions. We attempt to mimic this bias by creating simulated binaries and determining their combined magnitude before applying our speckle magnitude detection limits (see Section \ref{sec:methods}).

\section{Methods} \label{sec:methods}

To determine the number of binary companions in exoplanet host systems, we need to understand the fraction of stars and companions that can be detected by DSSI. We begin by simulating stellar populations such as those observed with \kk and DSSI using the population synthesis code \trilegal \citep{Girardi:aap2005a}, which simulates stellar parameters in any Galactic field. As noted previously, the \kk exoplanet sample is the result of community proposed targets with a range of stellar and exoplanet characteristics, making it nearly impossible to exactly replicate. We begin constructing our simulated sample by mimicking the distribution of \kk fields using star count simulations centered on the five \kk fields with significant numbers of observations (excluding only C1 and C3) and two additional pointings $\pm$ $5\degr$ from the center of each of the five fields. The locations of the resulting 15 \trilegal pointings are shown as plus signs in the upper plot of Figure \ref{fig:coords}. Each pointing has a field of view of one square degree. 

While \trilegal can produce binary stars with a given binary fraction and range of mass ratios, we simulate only single stars and manually add stellar companions to the \trilegal output so we have complete knowledge of their stellar and orbital properties. We begin by combining all of the simulated stars and applying cuts to mimic those made to our \kk observations. First, we create a distance-limited sample by omitting stars with $d > 1300$\,pc. We then select stars with effective temperatures between 3000\,K and 10,000\,K and surface gravities greater than $\log g > 3.0$. Because the focus of the \kepler/\kk missions is solar-type stars and we tend to focus on cooler stars in general, we also reduce the number of stars with $T_{\mathrm{eff}} > 7000$\,K by 50\%. After these cuts, the combined \trilegal sample contains over 66,000 stars. These cuts ensure that the simulated stars mimic our target sample and enable comparisons with other distance-limited samples of solar-type field stars as well as similar studies of \kepler stars. We applied the same distance, $T_{\mathrm{eff}}$, and $\log g$ cuts to our observed sample, as shown in Figure \ref{fig:parms}, though no stars with detected companions were eliminated (see Section \ref{subsec:stparms}).

\begin{figure}[t!]
\includegraphics[scale=0.48]{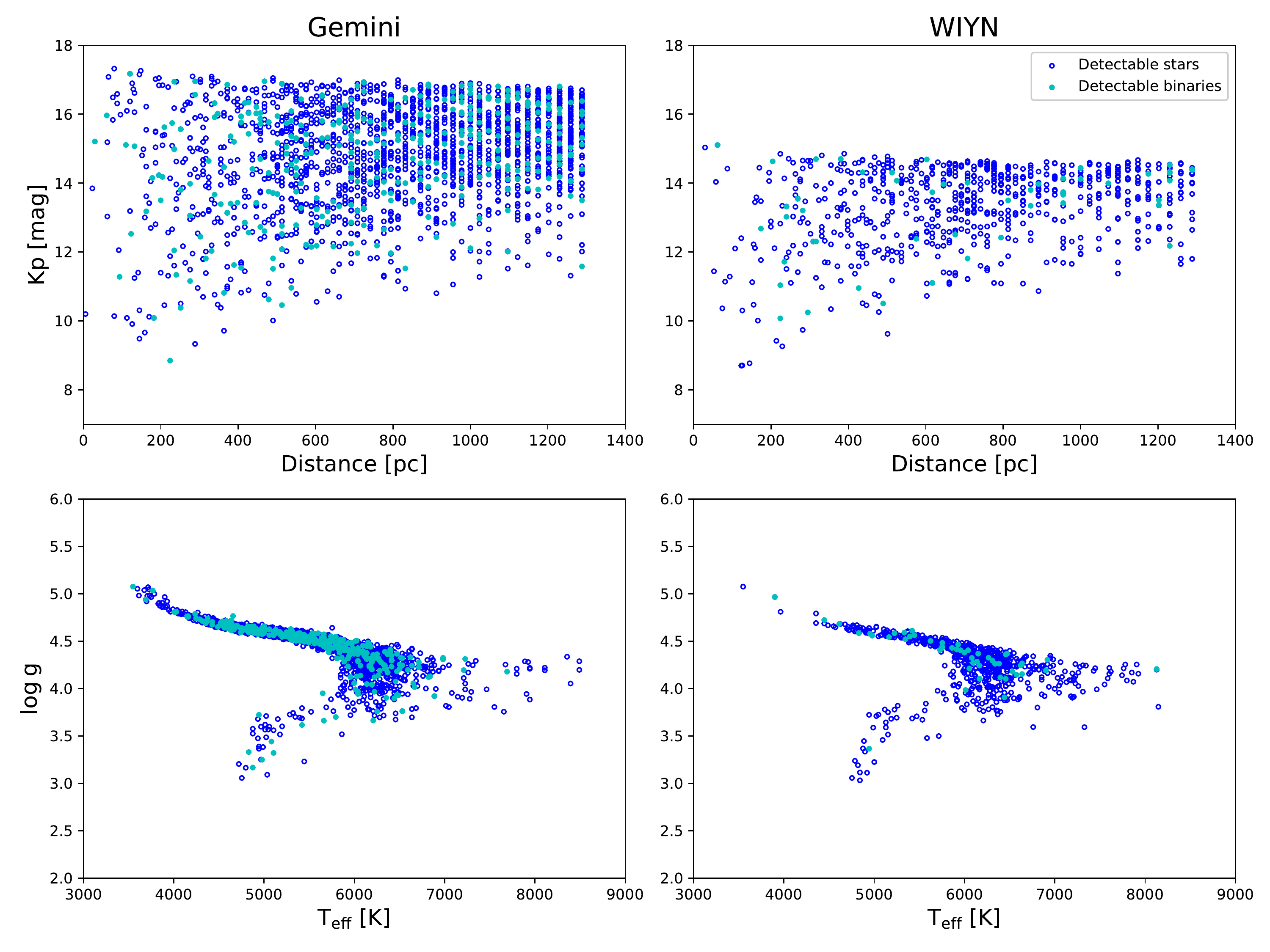}
\centering
\caption{Stellar properties for a randomly selected 20\% of the simulated stars from all 15 \trilegal pointings used in this work. Open blue circles represent stars that are detectable with DSSI using Gemini (left) or WIYN (right) but have no detectable companions, while filled cyan circles represent stars with detectable companions. The top row shows \kepler magnitude as a function of distance for the simulated stars, with surface gravity shown as a function of effective temperature in the bottom row. Stars with $d > 1300$\,pc and $\log g < 3.0$ have been excluded, and the number of stars with $T_{\mathrm{eff}} > 7000$\,K has been reduced by 50\% to better match the observed \kk stars shown in Figure \ref{fig:parms} (see Section \ref{sec:methods} for more details). \label{fig:TRIparms}}
\end{figure}

With the trimmed sample now distance-limited and composed of primarily solar-type (and some cooler) stars we add companions according to the statistics of solar-type field stars as determined by \citet{Duquennoy:aap1991a} and \citet{Raghavan:apjs2010a}. Based on \citeauthor{Raghavan:apjs2010a}, companion stars are added to the simulated population at a rate of 46\%. For each binary, the mass of the star output by \trilegal is used in combination with the mass ratio distribution of \citeauthor{Raghavan:apjs2010a} to determine the mass of the secondary. The mass of each companion is then used to determine its absolute $V$ magnitude via the mass-luminosity relationship of \citet{Henry:aj1993a}, which is converted to an apparent magnitude using the output \trilegal distance. The apparent $V$ magnitude of each component is then used to get a magnitude difference $\Delta m$, which is converted to the 880\,nm speckle filter by estimating the spectral type of each star based on its mass and applying the known filter transmission curve for DSSI.

We then determine orbital elements for each binary in order to determine position angles and separations at a randomly chosen observation epoch. Periods and eccentricities for each system were chosen according to the results of \citet{Duquennoy:aap1991a}, with a log-normal distribution of periods and eccentricities based on the period. We also assign random values of the inclination ($\cos i$), ascending node ($\Omega$), argument of periastron (angle between the node and periastron; $\omega$), and time of periastron passage ($T$) for each system. The masses and period for each binary are then used to calculate the semi-major axis in astronomical units (AU) and convert it into arcseconds using the distance from \trilegal. 

With the magnitude difference and separation of each component, we can now test for companions detectable using DSSI at Gemini or WIYN. To be observable the combined magnitude of the star and any companion must be brighter than the magnitude detection limit of $V = 14.5$ at WIYN or $V = 16.5$ at Gemini. As companions were added to the simulated stars before applying the magnitude detection limit, binaries where both components have magnitudes below the detection threshold but a combined magnitude above it will be detected. This reflects the bias for faint systems with small magnitude differences between the components inherent in magnitude-limited observations. Single stars above the magnitude limit are assumed to be seen as single by DSSI, but for binaries we apply a detection limit curve, which gives the maximum observable magnitude difference as a function of the separation from the primary star. Detection limit curves are determined for every reconstructed speckle image by measuring all local minima and maxima in concentric annuli around the central star and deriving the mean and standard deviation for each annulus. The detection limit for each annulus is then calculated as the average value of all maxima in the annulus plus five times the standard deviation of the maxima. For more details see \cite{Horch:aj2011a} and \citet{Howell:aj2011a}. If the magnitude difference of a companion is less than the value of the curve at the separation of the system then it will be detected. Here we use average detection limit curves (see Figure \ref{fig:detlim}) constructed from several unresolved objects that have apparent magnitudes between 11th and 14th magnitudes, comparable to the stars observed in this sample.


\begin{figure}[t!]
\includegraphics[scale=0.45]{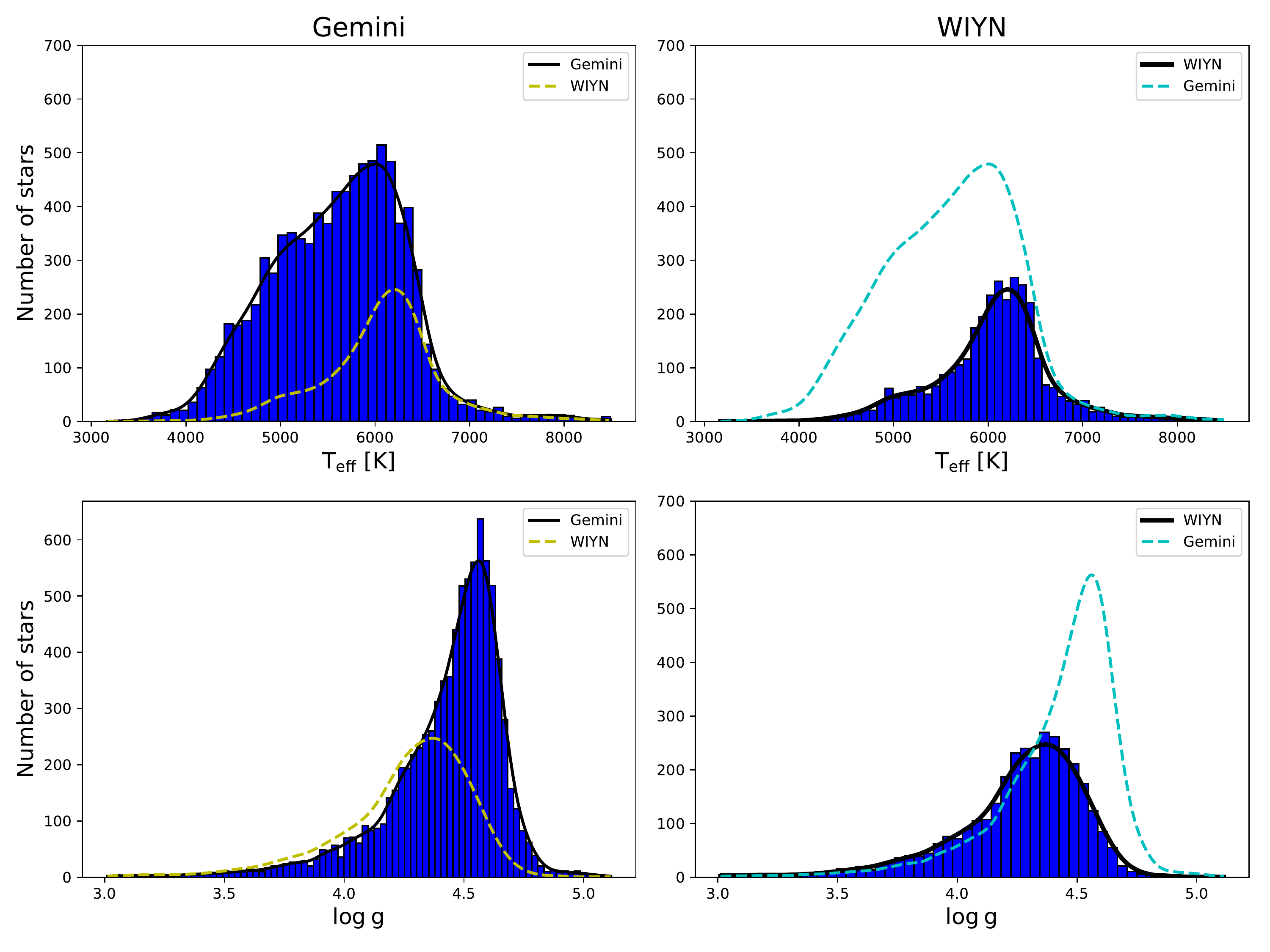}
\centering
\caption{Histograms of the entire sample of simulated stars observable with speckle imaging at Gemini (left) and WIYN (right). The distributions of effective temperatures (top) and surface gravity (bottom) are shown for both observatories, with a Gaussian kernel density estimate for each sample shown as a solid black line. For comparison, the scaled distributions from the other observatory (WIYN in yellow, Gemini in cyan) are shown on each plot, accentuating the subset of effective temperatures and surface gravities detectable from WIYN vs.~Gemini. \label{fig:TRIhist}}
\end{figure}

\begin{figure}[t!]
\includegraphics[scale=0.45]{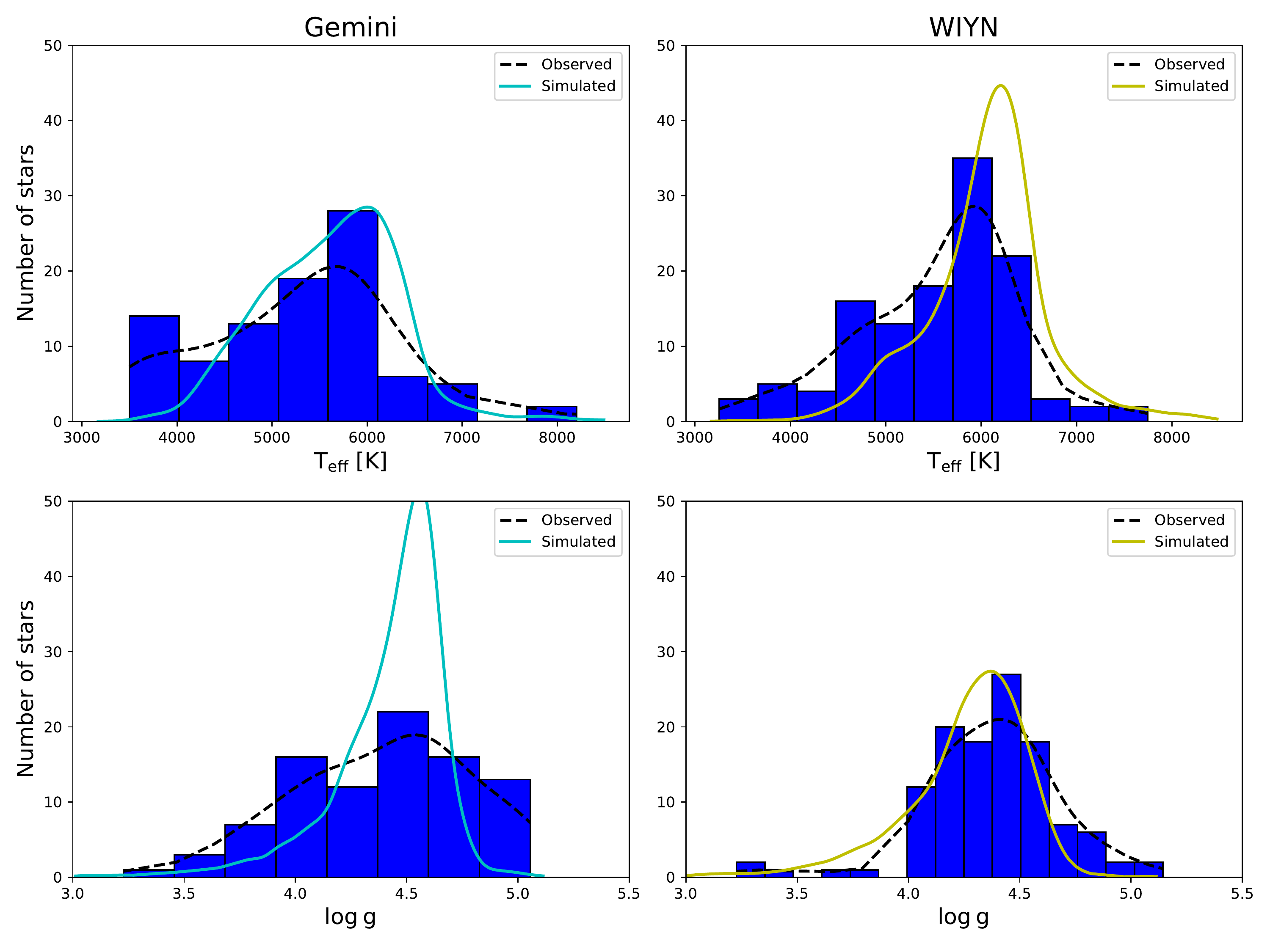}
\centering
\caption{Histograms of effective temperature (top) and surface gravity (bottom) for stars observed at Gemini (left) and WIYN (right). Normalized distributions using Gaussian kernel density estimates for the observed \kk and simulated \trilegal samples are over-plotted, with dashed black lines showing the observed distributions and solid cyan (Gemini) or yellow (WIYN) lines for the simulated distributions. The simulated distributions are generally similar to the observed distributions, but highlight our concentration on cooler stars, especially at Gemini. \label{fig:hist}}
\end{figure}

The results of the \trilegal simulations with the aforementioned cuts and detection limits applied are shown in Figure \ref{fig:TRIparms}. We plot a random 20\% of the approximately 8000 and 3000 stars detectable by Gemini and WIYN, respectively, in order to keep the figures legible. The simulated stars are plotted according to their stellar properties, with \kepler magnitude as a function of distance plotted in the top panels and surface gravity as a function of effective temperature on the bottom (comparable to Figure \ref{fig:parms}). Open blue circles represent stars that are detectable with DSSI but have no detectable companion, while filled cyan circles represent stars with detectable companions. 

In Figure \ref{fig:TRIhist} we show histograms of effective temperature and surface gravity for the entire sample of simulated stars detectable by DSSI at Gemini and WIYN. Gaussian kernel density estimates of each distribution are plotted as solid black lines. The scaled distributions from the opposite observatory (WIYN in yellow, Gemini in cyan) are over-plotted, demonstrating the larger range of stars observable at Gemini and the subset observable at WIYN. We also plot the simulated distribution against our observations in Figure \ref{fig:hist}. The solid cyan and yellow lines show the simulated distributions from Gemini and WIYN, respectively, while the dashed black lines show the Gaussian kernel density estimates of the observed distributions. The observed distributions are generally similar to the simulated distributions, but the smaller number of observed stars and emphasis on cooler systems with a broad range of $\log g$ is apparent, especially at Gemini.

In addition to physical binaries, DSSI also detects line-of-light companions or optical doubles. To determine the frequency of line-of-sight companions in each of the 15 \trilegal simulation fields, we assigned random positions within the field to each star and computed the on sky distance between these stars and those that are observable with DSSI (using the same distance, $T_{\mathrm{eff}}$, and $\log g$ cuts described above). For stars with separations less than 1.2 arsec, the magnitude difference between the components was computed and the average detection limit curves were used to determine whether they would be detected as doubles with DSSI.

\section{Results}\label{sec:results}
\subsection{Observed Companion Fractions}\label{subsec:obsfrac}

\begin{figure}[ht!]
\includegraphics[scale=0.58]{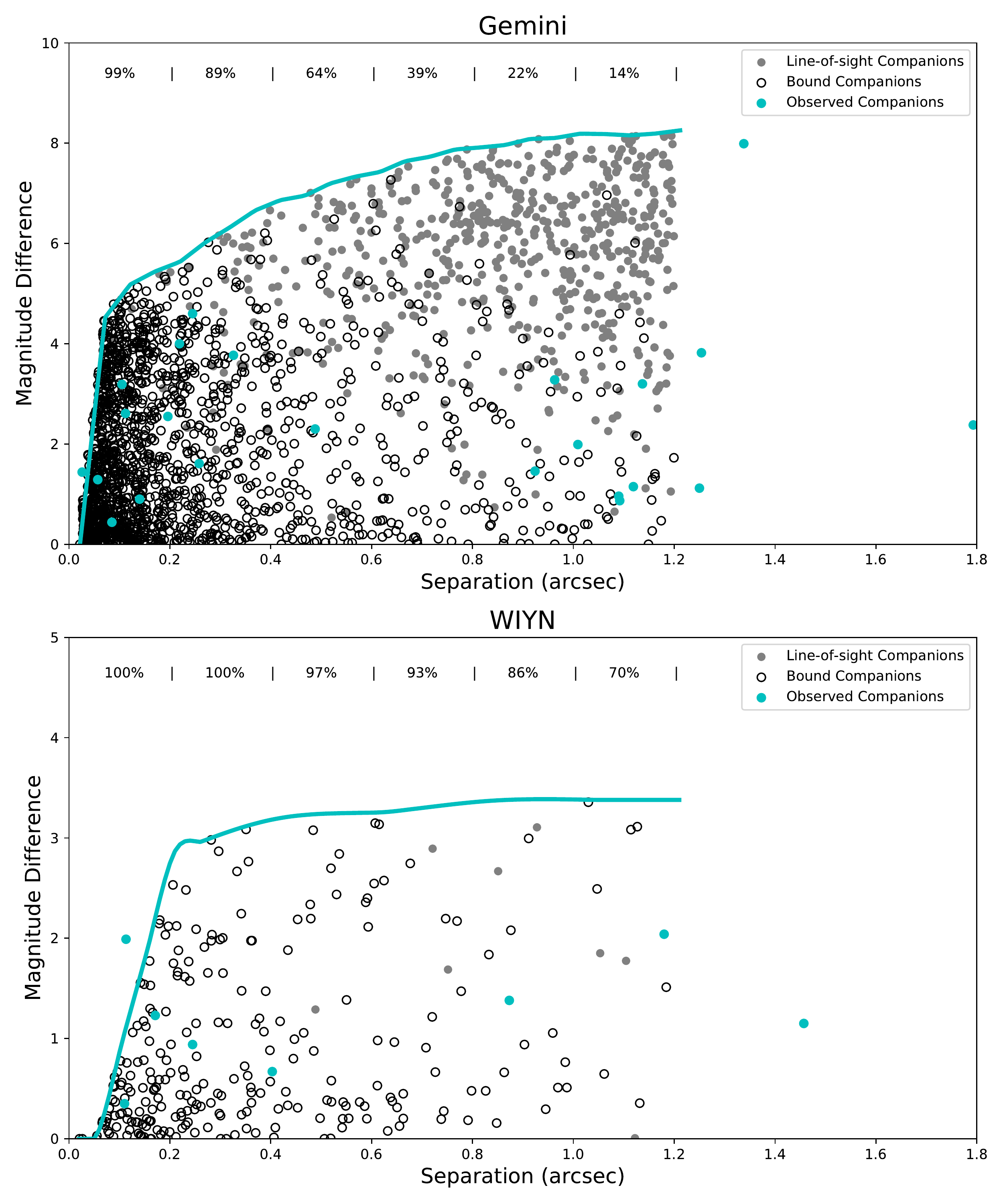}
\centering
\caption{Magnitude difference as a function of separation for real and simulated detections of stellar companions using DSSI. Open black circles represent simulated binaries with bound components detectable at Gemini (top) or WIYN (bottom), while filled gray circles represent detected line-of-sight companions from the simulations. Filled cyan circles show the separation and magnitude difference of stellar companions discovered with Gemini and WIYN. The cyan curve is an average detection limit curve for DSSI based on stars between 11th and 14th magnitudes. The numbers along the top of each plot give the percent of bound companions for each 0.2 arcsec wide bin in separation.\label{fig:detlim}}
\end{figure}

DSSI observations of \kk stars in C$0-$C7 detected 23 companions to 102 stars observed at Gemini and 8 companions to 127 stars observed at WIYN, resulting in raw companion fractions of $23\pm5$\,\% and $6\pm2$\,\%, respectively. Figure \ref{fig:detlim} shows the magnitude difference for the detected binaries as a function of separation, with the observed companions plotted as filled cyan circles. 
The average speckle detection limit curve is shown as a cyan line. Because the curve is an average of several detection limit curves for stars between 11th and 14th magnitudes, and such curves vary based on the magnitude of the stars and the seeing conditions, it is possible to detect companions with larger magnitude differences. One such detection is visible in each of the plots in Figure~\ref{fig:detlim}. Similarly, companions can be detected at separations larger than the end of the detection limit curve at 1.2 arcsec, however, the quality of the photometry ($\Delta m$) at such separations is not well quantified for DSSI as speckle patterns become less correlated at larger separations and speckles may fall out of the field when the separation is roughly equal to the instrument field of view. Therefore, to compare our observations with the simulated sample we exclude stars with observed companions that fall above or beyond the detection limit curves shown in Figure~\ref{fig:detlim}, resulting in observed companion fractions of $20\pm5\%$ (18 of 91 stars) at Gemini and $5\pm2\%$ (6 of 119 stars) at WIYN. 

\begin{deluxetable}{lCCCC}
\tablecaption{Simulated Companion Fractions \label{tab:campperc}}
\tablecolumns{5}
\tablenum{3}
\tablewidth{0pt}
\tablehead{
\colhead{} &
\multicolumn{2}{c}{Gemini} &
\multicolumn{2}{c}{WIYN} \\
\colhead{Location} &
\colhead{Bound} &
\colhead{Line-of-sight} &
\colhead{Bound} &
\colhead{Line-of-sight} 
}
\startdata
C0 - top      & 17\pm1\% & 4.2\pm0.6\% & 8\pm2 \%& 0.2\pm0.2\% \\
C0 - middle & 19\pm2\% & 3.9\pm0.6\% & 5\pm1 \%& 0.2\pm0.2\% \\
C0 - bottom & 18\pm1\% & 8.8\pm0.9\% &10\pm2 \%& 0.4\pm0.3\% \\
C4 - top      & 20\pm2\% & 1.2\pm0.4\% & 7\pm2 \%& 0\pm0\% \\
C4 - middle & 23\pm2\% & 0.7\pm0.4\% & 5\pm2 \%& 0\pm0\% \\
C4 - bottom & 21\pm2\% & 0.2\pm0.2\% & 10\pm3 \%& 0.5\pm0.5\% \\
C5 - top      & 17\pm3\% & 0\pm0\% & 7\pm3 \%& 0\pm0\% \\
C5 - middle & 21\pm3\% & 0.3\pm0.3\% & 11\pm3 \%& 0\pm0\% \\
C5 - bottom & 23\pm3\% & 0.5\pm0.3\% & 7\pm2 \%& 0\pm0\% \\
C6 - top       & 17\pm3\% & 0\pm0\% & 4\pm3 \%& 0\pm0\% \\
C6 - middle & 21\pm4\% & 0.4\pm0.4\% & 7\pm3 \%& 0\pm0\% \\
C6 - bottom & 20\pm3\% & 0.3\pm0.3\% & 7\pm3 \%& 0.8\pm0.8\% \\
C7 - top       & 18\pm1\% & 19\pm1\% & 7\pm1 \%& 0.2\pm0.2\% \\
C7 - middle & 19\pm1\% & 13\pm1\% & 7\pm1 \%& 0\pm0\% \\
C7 - bottom & 19\pm2\% & 11\pm1\% & 9\pm2 \%& 0.9\pm0.4\% \\
\hline
Total $\pm$ stdev & 19\pm2\% & 7\pm6\% & 8\pm2\% & 0.3\pm0.3\% \\
\enddata
\end{deluxetable}


\subsection{Simulated Companion Fractions}\label{subsec:simfrac}

The simulated results for stellar companions detectable with DSSI are also plotted in Figure~\ref{fig:detlim}, with bound companions plotted as open black circles and line-of-sight companions shown as filled gray circles. For Gemini, the percentage of detected companions predicted is $26\pm6\%$ (2100 out of 8787 detected stars), with $73\%$ predicted to be bound companions and the remaining $27\%$ optical doubles. At WIYN the predicted companion fraction is $8\pm2\%$ (261 out of 3337 detected stars), with $97\%$ predicted to be bound companions and $3\%$ optical doubles. The fractions of bound and line-of-sight companions determined for each of the 15 \trilegal pointings are listed in Table \ref{tab:campperc}. The overall predicted companion fraction for each observatory is determined by dividing the total number of binaries detectable in all 15 simulated fields by the total number of detectable stars in all fields. The uncertainties for the total bound and line-of-sight companion fractions are determined from the standard deviation of all 15 pointings, respectively (see Table~\ref{tab:campperc}), which were then added in quadrature to obtain the uncertainty in the overall companion fractions.

Table~\ref{tab:campperc} also highlights the relative crowding of different \kk campaign fields, as the fraction of optical doubles is highest in C0 and C7, which are near the Galactic Anti-center and Galactic center, respectively. The majority of line-of-sight companions are also found at larger separations and greater magnitude differences, as noted in \citet{Horch:apj2014a}, which is especially apparent in the Gemini simulation (see Figure~\ref{fig:detlim}). In contrast, the bound stellar companions cluster toward smaller separations and magnitude differences, implying that most sub-arcsec companions are gravitationally bound, especially in sparse fields. The percentage of bound companions for each 0.2 arsec bin in separation is listed along the top of Figure~\ref{fig:detlim}. 

\subsection{Comparing the Companion Fractions}\label{subsec:compare}

The results from the simulations (G: $26\pm6\%$, W: $8\pm2\%$) are similar to the observed companion fractions (G: $20\pm5\%$, W: $5\pm2\%$), and overlap within the uncertainties. This overlap increases if we include the companions observed beyond our average detection limits (points above the curve in Figure~\ref{fig:detlim}), raising the observed companion fractions to $21\pm5\%$ and $6\pm2\%$ for Gemini and WIYN, respectively. While stars with such magnitude differences are not allowed in the simulations, excluding these detections may underestimate our observed companion fraction as they likely balance out stars whose companions were not detected due to poor observing conditions, even though they would fall within our detection limits. This is especially true at WIYN, where the typical seeing is not as good as Gemini and we are often `pushing the limit' of observable magnitudes. 

The slightly higher companion fractions predicted by the simulated data could also indicate that the actual companion fraction for \kk stars is less than $46\%$. A lower companion fraction is possible as the sample of \kk stars includes M stars, unlike the sample of FGK field stars studied by \citet{Raghavan:apjs2010a}, and low-mass stars have fewer companions than solar-type stars \citep{Duchene:araa2013a, Winters:phd2015}. To determine the effect of a different input binary fraction, we varied the rate at which companions were added to the simulated stars and re-derived the bound companion fractions. For Gemini, the simulated companion fraction was equivalent to our observed rate using input binary fractions of $44\pm9\%$, while at WIYN we find an input binary fraction of $40\pm9\%$. These numbers are slightly lower than our initial binary fraction of $46\%$, but are equivalent within the uncertainties and consistent with the $40-50\%$ range for field stars.

As distance-limited samples can be subject to an overabundance of binaries based on underestimation of their photometric distances, we also re-examined our observed and simulated populations without the distance cuts at 1300\,pc. Including all of the stars in our observed sample, we have raw companion fractions of $23\pm5$\,\% and $6\pm2$\,\% for Gemini and WIYN, respectively, or $19\pm5$\,\% and $5\pm2$\,\% when eliminating the detections above and beyond the average detection limit curve plotted in Figure \ref{fig:detlim}, which are essentially identical to those with the distance cut. The \trilegal simulations, which are still trimmed in $\log g$ and $T_{\mathrm{eff}}$, are now constrained primarily by our limiting magnitudes at Gemini ($\lesssim$\,16.5) and WIYN ($\lesssim$\,14.5), which allow for simulated distances out to $\sim$\,7000\,pc and $\sim$\,5000\,pc, respectively. The resulting companion fraction for Gemini is $24\pm7$\,\% (6010 out of 25197 detected stars) and $8\pm3$\,\% (417 out of 5159 detected stars) for WIYN. While the total companion fractions are equivalent to the distance-limited simulations within the uncertainties, the fraction of line-of-sight companions observable from Gemini increases from $27\%$ to $42\%$ due to the larger number of stars included in the sample. If we then trim the distances to 1300\,pc (or 900\,pc to more closely match our observed K2 stars), we get companion fractions of 26\% (27\%), with 23\% line-of-sight companions, for Gemini and 10\% (11\%) for WIYN.

\section{Discussion}\label{sec:compare}

\subsection{Comparison with \kepler}
As noted earlier, \citet{Horch:apj2014a} conducted a similar study of exoplanet candidate host stars in the \kepler field of view, employing \trilegal simulations and DSSI detection limits to compare predicted and observed companion fractions at Gemini North and WIYN. They determined observed companion rates of $22.8\pm8.1\%$ (8 of 35 stars) at Gemini and $7.0\pm1.1\%$ (41 of 588 stars) at WIYN, which agreed within the uncertainties to their simulated results of $19.7\pm0.4\%$ and $7.8\pm0.4\%$, respectively. Their observed companion fractions are also consistent with what we observe for \kk (G: $20\pm5\%$, W: $5\pm2\%$). The predicted companion fraction for WIYN is also the same between the two studies, however, we derive a larger companion fraction at Gemini. This is likely due to the more crowded fields observed with \kk (specifically C0 and C7), as we derive a higher percentage of optical doubles at Gemini ($27\%$ vs.~their $16\%$). 

Unlike with \kepler, however, the \kk planet candidates have not been as strictly vetted and there is often conflicting evidence on whether a planet remains a candidate or is a false positive. As such, we cannot produce a clean sample of confirmed planet host stars in order to make a conclusive statement about their binarity. If we only consider planet candidates with no data suggesting they are false positives or eclipsing binaries on the ExoFOP website, the observed companion fraction drops to $10\%$ for stars observed with Gemini, but remains at $5\%$ for WIYN. However, none of the stars omitted as potential false positives have follow-up work that excludes the existence of a planet, and the occurrence rate statistics of \kepler suggest most stars (and wide binaries) are likely to have at least one planet \citep{Burke:apj2015a}. Therefore, while we cannot state that all of the stars observed have validated planets, we can conclude that the companion fraction of $40-50\%$ found in nearby field stars \citep{Raghavan:apjs2010a} and the \kepler field \citep{Horch:apj2014a} is consistent with the \kk Campaign fields.

\subsection{Close Stellar Companions and Exoplanets}

\begin{figure}[h!]
\includegraphics[scale=0.5]{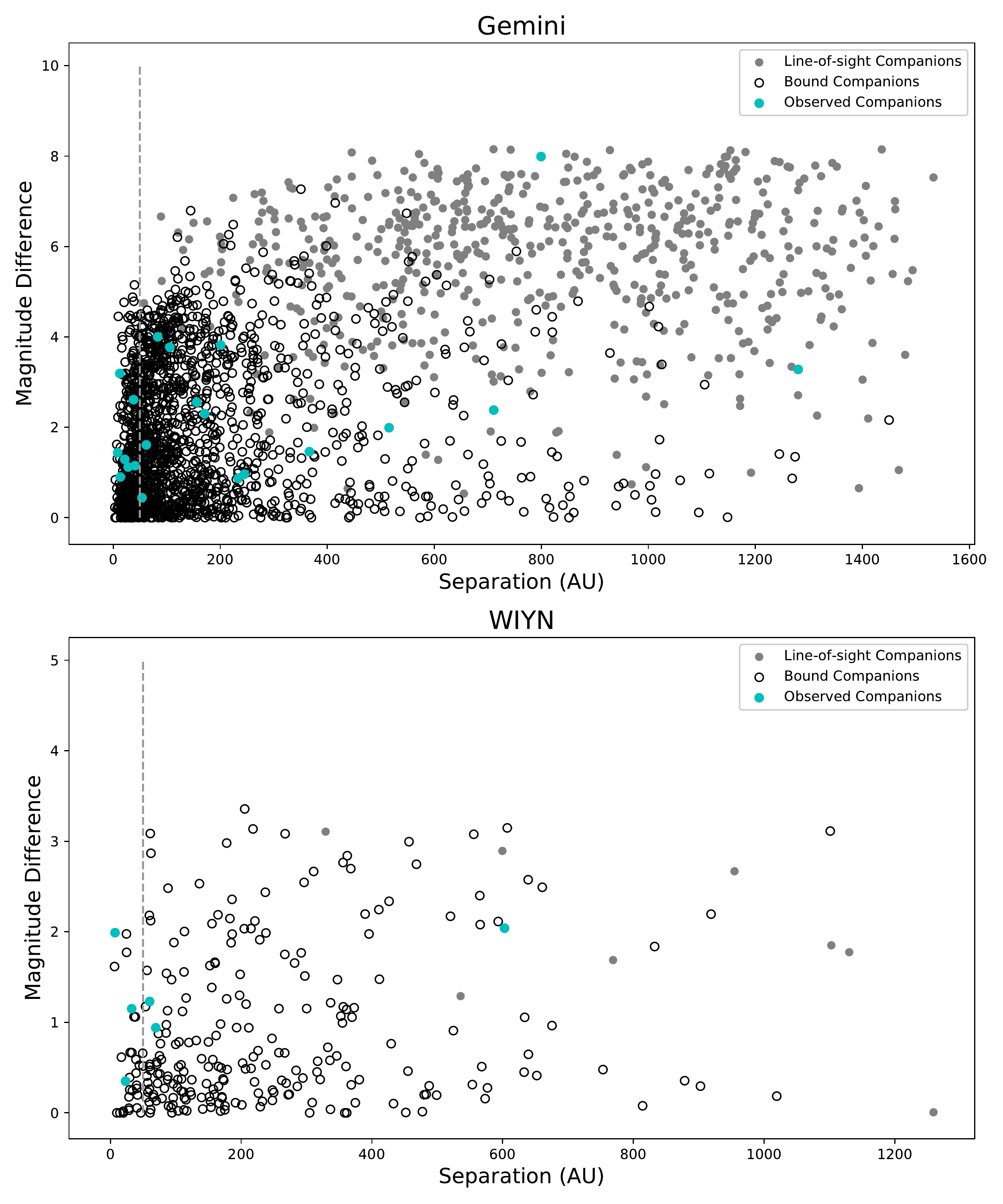}
\centering
\caption{Real and simulated detections of stellar companions plotted as a function of separation in AU. Open black circles represent simulated binaries with bound components detectable at Gemini (top) or WIYN (bottom), while filled gray circles represent detected line-of-sight companions from the simulations. Filled cyan circles show the separation and magnitude difference of stellar companions discovered with Gemini and WIYN. A gray dashed line is plotted at 50\,AU, the semi-major axis within which stellar companions are suppressed in exoplanet systems according to \citet{Kraus:aj2016a}.
\label{fig:ausep}}
\end{figure}

Several studies have suggested that close binaries are suppressed among populations of exoplanet host stars. \cite{Horch:apj2014a} and \citet{Deacon:mnras2016a} looked at the rates of wide binaries among exoplanet host stars in \kepler and found no difference from the binary rate among field stars. However, \citet{Wang:apj2014b} and \citet{Kraus:aj2016a} have concluded that stellar companions are suppressed inside $\sim20-50$\,AU for exoplanet host stars. \citet{Wang:apj2014b} examined a sample of 138 multi-planet candidate systems from \kepler and found a stellar multiplicity rate significantly lower than field stars for semi-major axes less than 20\,AU, and possible suppression out to 85\,AU. \cite{Kraus:aj2016a} used high-resolution imaging to detect 506 companions to 382 \kepler planet candidates, combining adaptive optics and aperture masking to probe for companions on solar-system scales of $1.5-50$\,AU. They found that close binary companions within $\sim$50\,AU are suppressed by a factor of 34\% in systems with exoplanets, implying that it is rare for planetary systems to form when a binary companion is present on solar-system scales.

\begin{figure}[t!]
\includegraphics[scale=0.6]{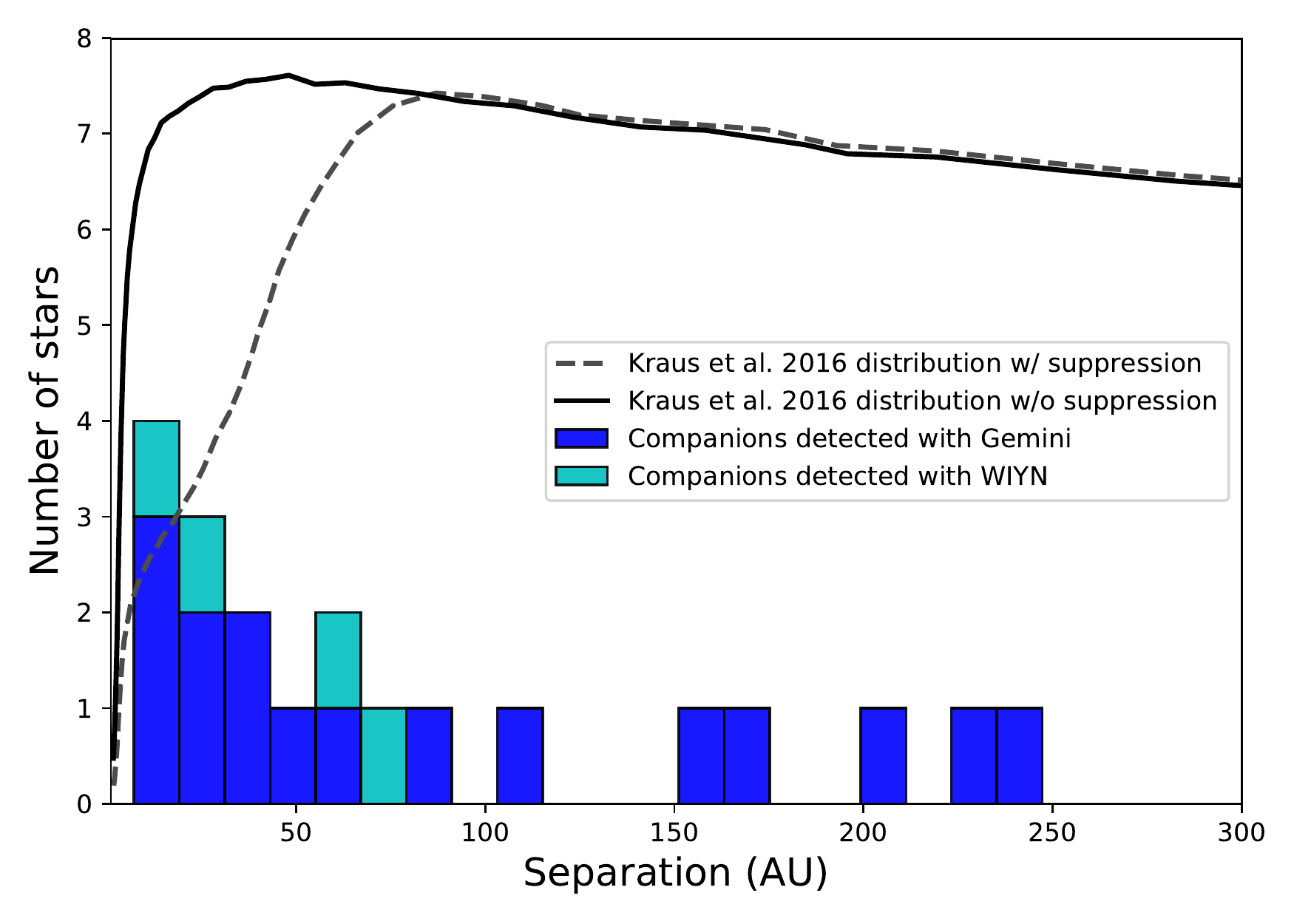}
\centering
\caption{Histogram showing observed companion separations within 300\,AU for Gemini (blue) and WIYN (cyan). Projected separation distributions from \citet{Kraus:aj2016a} have been scaled to our data and over-plotted. The solid black line shows their predicted distribution of binary companions with mass ratios larger than 0.4 based on \citet{Raghavan:apjs2010a}, while the dashed line shows the same distribution with binary companions suppressed by a factor of 0.34 inside of 47\,AU (see Figures 7 and 8 in \citeauthor{Kraus:aj2016a}). While our sample size is too small to draw any definitive conclusions regarding the suppression of close stellar companions, setting the bin size to a (unreasonable) value of 12\,AU shows no apparent suppression of stellar companions within $\sim$\,50\,AU and our distribution remains peaked toward smaller separations.
\label{fig:auhist}}
\end{figure}

With DSSI's ability to detect companion stars at the diffraction limit of the telescope ($\sim 0.05"$ at WIYN, $\sim 0.027"$ at Gemini) and the closer distances of many of the \kk stars, we can examine our sample of planet candidate host stars for evidence of  suppression of close binary companions. Of the 29 companions detected with speckle imaging, 26 of them have distance estimates provided in the EPIC that we can use to convert the observed separations from arseconds to AU. The magnitude difference of the real and simulated stellar companions are plotted as a function of separation in AU in Figure~\ref{fig:ausep}. The gray dashed line marks the $\sim50$\,AU semi-major axis cutoff found by \cite{Kraus:aj2016a}. Note that none of the simulated line-of-sight companions fall in this region. At Gemini, 8 of the 21 ($38\%$) observed companions are within 50\,AU of the primary star, while 3 of the 6 ($50\%$) companions detected at WIYN are within 50\,AU (one companion was detected by both telescopes). Figure~\ref{fig:auhist} shows a histogram of the \kk companions detected with Gemini and WIYN as a function of separation out to 300\,AU, as well as projected separation distributions from \citet{Kraus:aj2016a} with (dashed line) and without (solid black line) companion suppression. The distributions from their Figures 7 and 8 were scaled down to our sample size of 26 and are plotted linearly. Using a (unreasonable) bin size of 12\,AU to examine the companions observed inside of 50\,AU, we do not see evidence of companion suppression as our distribution is peaked towards smaller separations and more generally follows the projected separation curve without suppression.

As \kk includes more M-stars then the \kepler sample, and smaller stars have planets at smaller separations relative to solar-type stars, we examined the spectral types of the stars that have companions within $\sim$\,50\,AU to ensure we are not looking exclusively at companions to smaller stars. However, only 3 of the 10 stars 
with close stellar companions are M-dwarfs, and none fall in the first bin of Figure \ref{fig:hist}. 
Therefore, while we cannot make any definitive statements regarding the companion fraction of exoplanet host stars at close separations due to the small number of stellar companions observed to date and the unconfirmed nature of several of the planets, our observations do not indicate a lower rate of close-in stellar companions.

\section{Summary}\label{sec:conc} 

In this work we compare the observed companion fraction for a sample of 206 exoplanet host stars detected by \kk and observed with DSSI speckle imaging to simulated stars for five different \kk campaign fields. The stars were simulated using \trilegal, with stellar companions added according to \citet{Raghavan:apjs2010a}, and the detectable companions determined by applying the established detection limits of speckle imaging. The observed companion rates of $20\pm5\%$ using Gemini and $5\pm2$\% using WIYN agree within uncertainties to the expected rates of detected companions of $26\pm6\%$ and $8\pm2$\% for Gemini and WIYN, respectively. Therefore, based on the expected binary fraction of $46\%$ from \citet{Raghavan:apjs2010a}, we show that the \kk exoplanet host stars have a binary fraction of $40-50\%$, similar to that of field stars and \kepler exoplanet host stars. 

The simulations also predict that a majority of stellar companions within $\sim$\,$1''$ will be bound to the host star, though the magnitude difference and congestion of the field should be considered (See Figure~\ref{fig:detlim}). Finally, using the magnitudes and separations probed by DSSI speckle observations we do not see evidence of stellar companion suppression within $\sim$\,$50$\,AU of exoplanet host stars. However, more high-angular resolution images are needed to probe solar-system scales around additional exoplanet hosts. Accordingly, we are using our upgraded speckle imagers, NESSI at WIYN \citep{Scott:pasp2018a} and 'Alopeke \citep{Scott:2018b} at Gemini North, to continue observing the bright and close \kk exoplanet hosts for which our resolution limits extend to such scales.

\acknowledgments

The authors would like to thank all of the excellent staff at the WIYN and Gemini telescopes for their help during our observing runs. \kepler was competitively selected as the tenth Discovery mission. Funding for this mission is provided by NASA's Science Mission Directorate. We also gratefully acknowledge the role of the \kepler Science Office in upgrading DSSI to the two-EMCCD mode. This research has made use of the NASA Exoplanet Archive and the Exoplanet Follow-up Observation Program website, which are operated by the California Institute of Technology, under contract with NASA under the Exoplanet Exploration Program. RM's research was supported by an appointment to the NASA Postdoctoral Program at the NASA Ames Research Center, administered by Universities Space Research Association under contract with NASA.

\vspace{5mm}

\facilities{WIYN, Gemini:Gillett, Gemini:South}

\software{TRILEGAL \citep{Girardi:2015a}}

\clearpage

\bibliography{../../SpeckleLIB}

\begin{thebibliography}{}
\expandafter\ifx\csname natexlab\endcsname\relax\def\natexlab#1{#1}\fi
\providecommand{\url}[1]{\href{#1}{#1}}

\bibitem[{{Brown}(2003)}]{Brown:apjl2003a}
{Brown}, T.~M. 2003, \apjl, 593, L125

\bibitem[{{Burke} {et~al.}(2015){Burke}, {Christiansen}, {Mullally}, {Seader},
  {Huber}, {Rowe}, {Coughlin}, {Thompson}, {Catanzarite}, {Clarke}, {Morton},
  {Caldwell}, {Bryson}, {Haas}, {Batalha}, {Jenkins}, {Tenenbaum}, {Twicken},
  {Li}, {Quintana}, {Barclay}, {Henze}, {Borucki}, {Howell}, \&
  {Still}}]{Burke:apj2015a}
{Burke}, C.~J., {Christiansen}, J.~L., {Mullally}, F., {et~al.} 2015, \apj,
  809, 8

\bibitem[{{Ciardi} {et~al.}(2015){Ciardi}, {Beichman}, {Horch}, \&
  {Howell}}]{Ciardi:apj2015a}
{Ciardi}, D.~R., {Beichman}, C.~A., {Horch}, E.~P., \& {Howell}, S.~B. 2015,
  \apj, 805, 16

\bibitem[{{Crossfield} {et~al.}(2016){Crossfield}, {Ciardi}, {Petigura},
  {Sinukoff}, {Schlieder}, {Howard}, {Beichman}, {Isaacson}, {Dressing},
  {Christiansen}, {Fulton}, {L{\'e}pine}, {Weiss}, {Hirsch}, {Livingston},
  {Baranec}, {Law}, {Riddle}, {Ziegler}, {Howell}, {Horch}, {Everett}, {Teske},
  {Martinez}, {Obermeier}, {Benneke}, {Scott}, {Deacon}, {Aller}, {Hansen},
  {Mancini}, {Ciceri}, {Brahm}, {Jord{\'a}n}, {Knutson}, {Henning}, {Bonnefoy},
  {Liu}, {Crepp}, {Lothringer}, {Hinz}, {Bailey}, {Skemer}, \&
  {Defrere}}]{Crossfield:apjs2016a}
{Crossfield}, I.~J.~M., {Ciardi}, D.~R., {Petigura}, E.~A., {et~al.} 2016,
  \apjs, 226, 7

\bibitem[{{Deacon} {et~al.}(2016){Deacon}, {Kraus}, {Mann}, {Magnier},
  {Chambers}, {Wainscoat}, {Tonry}, {Kaiser}, {Waters}, {Flewelling}, {Hodapp},
  \& {Burgett}}]{Deacon:mnras2016a}
{Deacon}, N.~R., {Kraus}, A.~L., {Mann}, A.~W., {et~al.} 2016, \mnras, 455,
  4212

\bibitem[{{Dressing} {et~al.}(2017){Dressing}, {Vanderburg}, {Schlieder},
  {Crossfield}, {Knutson}, {Newton}, {Ciardi}, {Fulton}, {Gonzales}, {Howard},
  {Isaacson}, {Livingston}, {Petigura}, {Sinukoff}, {Everett}, {Horch}, \&
  {Howell}}]{Dressing:aj2017a}
{Dressing}, C.~D., {Vanderburg}, A., {Schlieder}, J.~E., {et~al.} 2017, \aj,
  154, 207

\bibitem[{{Duch{\^e}ne} \& {Kraus}(2013)}]{Duchene:araa2013a}
{Duch{\^e}ne}, G., \& {Kraus}, A. 2013, \araa, 51, 269

\bibitem[{{Duquennoy} \& {Mayor}(1991)}]{Duquennoy:aap1991a}
{Duquennoy}, A., \& {Mayor}, M. 1991, \aap, 248, 485

\bibitem[{{Everett} {et~al.}(2015){Everett}, {Barclay}, {Ciardi}, {Horch},
  {Howell}, {Crepp}, \& {Silva}}]{Everett:aj2015a}
{Everett}, M.~E., {Barclay}, T., {Ciardi}, D.~R., {et~al.} 2015, \aj, 149, 55

\bibitem[{{Furlan} \& {Howell}(2017)}]{Furlan:aj2017b}
{Furlan}, E., \& {Howell}, S.~B. 2017, \aj, 154, 66

\bibitem[{{Furlan} {et~al.}(2017){Furlan}, {Ciardi}, {Everett}, {Saylors},
  {Teske}, {Horch}, {Howell}, {van Belle}, {Hirsch}, {Gautier}, {Adams},
  {Barrado}, {Cartier}, {Dressing}, {Dupree}, {Gilliland}, {Lillo-Box},
  {Lucas}, \& {Wang}}]{Furlan:aj2017a}
{Furlan}, E., {Ciardi}, D.~R., {Everett}, M.~E., {et~al.} 2017, \aj, 153, 71

\bibitem[{{Girardi} {et~al.}(2015){Girardi}, {Barbieri}, {Miglio}, {Bossini},
  {Bressan}, {Marigo}, \& {Rodrigues}}]{Girardi:2015a}
{Girardi}, L., {Barbieri}, M., {Miglio}, A., {et~al.} 2015, in Astrophysics and
  Space Science Proceedings, Vol.~39, Asteroseismology of Stellar Populations
  in the Milky Way, ed. A.~{Miglio}, P.~{Eggenberger}, L.~{Girardi}, \&
  J.~{Montalb{\'a}n}, 125

\bibitem[{{Girardi} {et~al.}(2005){Girardi}, {Groenewegen}, {Hatziminaoglou},
  \& {da Costa}}]{Girardi:aap2005a}
{Girardi}, L., {Groenewegen}, M.~A.~T., {Hatziminaoglou}, E., \& {da Costa}, L.
  2005, \aap, 436, 895

\bibitem[{{Gray}(2008)}]{Gray:2008a}
{Gray}, D.~F. 2008, {The Observation and Analysis of Stellar Photospheres}
  (Cambridge, UK: Cambridge University Press)

\bibitem[{{Haghighipour}(2006)}]{Haghighipour:apj2006a}
{Haghighipour}, N. 2006, \apj, 644, 543

\bibitem[{{Henry} \& {McCarthy}(1993)}]{Henry:aj1993a}
{Henry}, T.~J., \& {McCarthy}, Jr., D.~W. 1993, \aj, 106, 773

\bibitem[{{Hirsch} {et~al.}(2017){Hirsch}, {Ciardi}, {Howard}, {Everett},
  {Furlan}, {Saylors}, {Horch}, {Howell}, {Teske}, \& {Marcy}}]{Hirsch:aj2017a}
{Hirsch}, L.~A., {Ciardi}, D.~R., {Howard}, A.~W., {et~al.} 2017, \aj, 153, 117

\bibitem[{{Horch} {et~al.}(2011{\natexlab{a}}){Horch}, {Gomez}, {Sherry},
  {Howell}, {Ciardi}, {Anderson}, \& {van Altena}}]{Horch:aj2011b}
{Horch}, E.~P., {Gomez}, S.~C., {Sherry}, W.~H., {et~al.} 2011{\natexlab{a}},
  \aj, 141, 45

\bibitem[{{Horch} {et~al.}(2012){Horch}, {Howell}, {Everett}, \&
  {Ciardi}}]{Horch:aj2012a}
{Horch}, E.~P., {Howell}, S.~B., {Everett}, M.~E., \& {Ciardi}, D.~R. 2012,
  \aj, 144, 165

\bibitem[{{Horch} {et~al.}(2014){Horch}, {Howell}, {Everett}, \&
  {Ciardi}}]{Horch:apj2014a}
---. 2014, \apj, 795, 60

\bibitem[{{Horch} {et~al.}(2011{\natexlab{b}}){Horch}, {van Altena}, {Howell},
  {Sherry}, \& {Ciardi}}]{Horch:aj2011a}
{Horch}, E.~P., {van Altena}, W.~F., {Howell}, S.~B., {Sherry}, W.~H., \&
  {Ciardi}, D.~R. 2011{\natexlab{b}}, \aj, 141, 180

\bibitem[{{Horch} {et~al.}(2009){Horch}, {Veillette}, {Baena Gall{\'e}},
  {Shah}, {O'Rielly}, \& {van Altena}}]{Horch:aj2009a}
{Horch}, E.~P., {Veillette}, D.~R., {Baena Gall{\'e}}, R., {et~al.} 2009, \aj,
  137, 5057

\bibitem[{{Howell} {et~al.}(2016){Howell}, {Everett}, {Horch}, {Winters},
  {Hirsch}, {Nusdeo}, \& {Scott}}]{Howell:apjl2016a}
{Howell}, S.~B., {Everett}, M.~E., {Horch}, E.~P., {et~al.} 2016, \apjl, 829,
  L2

\bibitem[{{Howell} {et~al.}(2011){Howell}, {Everett}, {Sherry}, {Horch}, \&
  {Ciardi}}]{Howell:aj2011a}
{Howell}, S.~B., {Everett}, M.~E., {Sherry}, W., {Horch}, E., \& {Ciardi},
  D.~R. 2011, \aj, 142, 19

\bibitem[{{Howell} {et~al.}(2014){Howell}, {Sobeck}, {Haas}, {Still},
  {Barclay}, {Mullally}, {Troeltzsch}, {Aigrain}, {Bryson}, {Caldwell},
  {Chaplin}, {Cochran}, {Huber}, {Marcy}, {Miglio}, {Najita}, {Smith},
  {Twicken}, \& {Fortney}}]{Howell:pasp2014a}
{Howell}, S.~B., {Sobeck}, C., {Haas}, M., {et~al.} 2014, \pasp, 126, 398

\bibitem[{{Huber} {et~al.}(2016){Huber}, {Bryson}, {Haas}, {Barclay},
  {Barentsen}, {Howell}, {Sharma}, {Stello}, \& {Thompson}}]{Huber:apjs2016a}
{Huber}, D., {Bryson}, S.~T., {Haas}, M.~R., {et~al.} 2016, \apjs, 224, 2

\bibitem[{{Jang-Condell}(2015)}]{Jang-Condell:apj2015a}
{Jang-Condell}, H. 2015, \apj, 799, 147

\bibitem[{{Kraus} {et~al.}(2016){Kraus}, {Ireland}, {Huber}, {Mann}, \&
  {Dupuy}}]{Kraus:aj2016a}
{Kraus}, A.~L., {Ireland}, M.~J., {Huber}, D., {Mann}, A.~W., \& {Dupuy}, T.~J.
  2016, \aj, 152, 8

\bibitem[{{Rafikov} \& {Silsbee}(2015{\natexlab{a}})}]{Rafikov:apj2015a}
{Rafikov}, R.~R., \& {Silsbee}, K. 2015{\natexlab{a}}, \apj, 798, 69

\bibitem[{{Rafikov} \& {Silsbee}(2015{\natexlab{b}})}]{Rafikov:apj2015b}
---. 2015{\natexlab{b}}, \apj, 798, 70

\bibitem[{{Raghavan} {et~al.}(2010){Raghavan}, {McAlister}, {Henry}, {Latham},
  {Marcy}, {Mason}, {Gies}, {White}, \& {ten Brummelaar}}]{Raghavan:apjs2010a}
{Raghavan}, D., {McAlister}, H.~A., {Henry}, T.~J., {et~al.} 2010, \apjs, 190,
  1

\bibitem[{{Scott} {et~al.}(2018{\natexlab{a}}){Scott}, {Howell}, {Horch}, \&
  {Everett}}]{Scott:pasp2018a}
{Scott}, N.~J., {Howell}, S.~B., {Horch}, E.~P., \& {Everett}, M.~E.
  2018{\natexlab{a}}, \pasp, 130, 054502

\bibitem[{{Scott} {et~al.}(2018{\natexlab{b}}){Scott}, {Howell}, {Horch}, \&
  {Everett}}]{Scott:2018b}
---. 2018{\natexlab{b}}, in prep.

\bibitem[{{Thebault} \& {Haghighipour}(2015)}]{Thebault:2015a}
{Thebault}, P., \& {Haghighipour}, N. 2015, {Planet Formation in Binaries}, ed.
  S.~{Jin}, N.~{Haghighipour}, \& W.-H. {Ip} (Springer-Verlag Berlin
  Heidelberg), 309--340

\bibitem[{{Vanderburg} {et~al.}(2016){Vanderburg}, {Latham}, {Buchhave},
  {Bieryla}, {Berlind}, {Calkins}, {Esquerdo}, {Welsh}, \&
  {Johnson}}]{Vanderburg:apjs2016a}
{Vanderburg}, A., {Latham}, D.~W., {Buchhave}, L.~A., {et~al.} 2016, \apjs,
  222, 14

\bibitem[{{Wang} {et~al.}(2014{\natexlab{a}}){Wang}, {Fischer}, {Xie}, \&
  {Ciardi}}]{Wang:apj2014b}
{Wang}, J., {Fischer}, D.~A., {Xie}, J.-W., \& {Ciardi}, D.~R.
  2014{\natexlab{a}}, \apj, 791, 111

\bibitem[{{Wang} {et~al.}(2014{\natexlab{b}}){Wang}, {Xie}, {Barclay}, \&
  {Fischer}}]{Wang:apj2014a}
{Wang}, J., {Xie}, J.-W., {Barclay}, T., \& {Fischer}, D.~A.
  2014{\natexlab{b}}, \apj, 783, 4

\bibitem[{{Winters}(2015)}]{Winters:phd2015}
{Winters}, J.~G. 2015, PhD thesis, Georgia State University, Atlanta, GA

\end{thebibliography}

\end{document}